\documentstyle[aps,prd]{revtex} 

\input psfig.sty
\newcommand{\be}{\begin{equation}}
\newcommand{\ba}{\begin{eqnarray}}
\newcommand{\ee}{\end{equation}}
\newcommand{\ea}{\end{eqnarray}}

\begin{document}
\draft

\title{Chaos and order in a finite universe}
\author{John D. Barrow${}^*$ and Janna Levin${}^{**}$}
\address{${}^{*}$DAMTP, Cambridge University,
Silver St., Cambridge CB3 9EW
\quad (J.D.Barrow@damtp.cam.ac.uk)}
\address{${}^{**}$Astronomy Centre, University of Sussex,
Brighton BN1 9QJ, UK \quad
(janna@astr.cpes.susx.ac.uk)}

\twocolumn[\hsize\textwidth\columnwidth\hsize\csname
           @twocolumnfalse\endcsname

\maketitle
\widetext

\begin{abstract}

All inhabitants
of this  universe, from galaxies to people, are finite.  Yet the universe
itself is often assumed to be infinite.  
If instead the universe is topologically finite, then light and matter 
can take chaotic paths around the compact geometry.
Chaos may lead to ordered features in 
the distribution of matter throughout space.

\end{abstract}
\bigskip
\medskip
\centerline{Contribution to the conference proceedings for ``The Chaotic
Universe'', ICRA, Rome.}
\bigskip
\medskip
]

\narrowtext
\setcounter{section}{1}

In cosmology as well as string theory, compact spaces have received
renewed attention.  Most discussions evade the chaos inherent in many
of these spaces.  Here we pursue the consequences of chaos on a
compact hyperbolic space by isolating the fractal set of closed loop
orbits.
We also discuss the implications this may have for the distribution 
of large-scale structure in our own cosmos.

Compact hyperbolic spaces are known to induce chaotic mixing of
trajectories as they wrap around the space.  The closed loop orbits,
though seemingly
special, define the entire structure of the chaotic dynamics.  The
dense and abundant periodic orbits pack themselves into the finite space
by collectively forming a fractal. 

\begin{figure}[tbp]
\centerline{{
\psfig{file=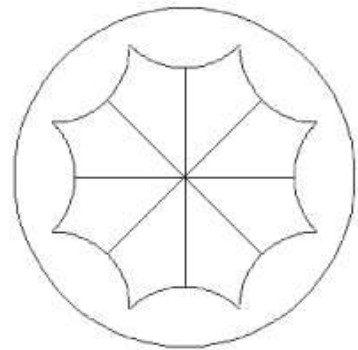,width=2.0in}}} \vskip 15truept
\caption{The fundamental octagon, drawn above in the Poincar\'e unit
sphere, is made finite by gluing opposite faces.  The shortest closed 
orbits repeatedly exit one face and enter the opposite face.
}
\label{firstorb}
\end{figure}

For simplicity we view the closed loop
null-geodesics on a $2D$ finite space.
Consider the double donut built by cutting a regular octagon out of a
hyperbolic $2D$ space and identifying the opposite sides in pairs.
The fundamental domain in fig.\ref{firstorb} is drawn on the Poincar\'e unit
sphere with the metric
	\be
	ds^2=-d\eta^2+{4\over (1-r^2)^2}(dr^2+r^2d\phi^2) \ \ .
	\ee
Geodesics are semi-circles which are orthogonal to the boundary at $r=1$.  
The shortest closed loop orbits are also drawn in fig. \ref{firstorb}.
The
null geodesics are completely specified by the angular momentum 
$L=4(1-r^2)^{-2}r^2\dot\phi$ and the angular coordinate 
$\theta $ on the
boundary at which the geodesics originated.
As geodesics exit and re-enter the fundamental domain, they are chaotically
mixed.  A re-entry map can be found given the rules for identifying the faces
of the octagon \cite{bv}.  The closed loop orbits can be found systematically
\cite{lb} order by order in the number of windings around the space.

\begin{figure}[tbp]
\centerline{{
\psfig{file=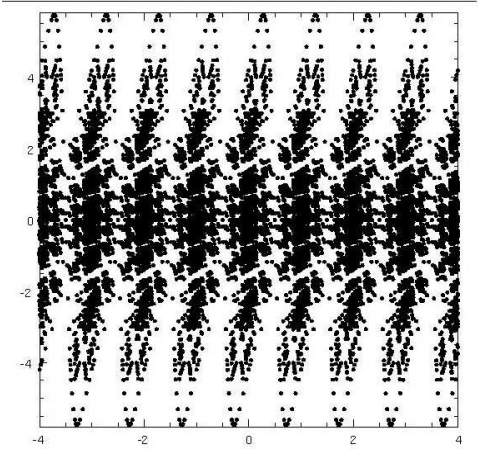,width=2.5in}}} \vskip 15truept
\caption{The fractal set of periodic orbits in the $(\theta,L)$ plane
shown to order $n=5$.  The $x$-axis corresponds to $\theta $ in units
of $\pi/4$.
}
\label{fractal}
\end{figure}

In fig. \ref{fractal}, all of the periodic orbits are shown which execute $5$
windings or less around the octagon.  There are $19,624$ such orbits.
We find the box counting dimension of the set by covering it with boxes of size
$\epsilon$ on a side and counting the growth in the number of boxes needed to
cover the set as $\epsilon$ gets smaller.  The dimension is
found to be
$D_0=\lim_{\epsilon\rightarrow 0}\ln N(\epsilon)/\ln(1/\epsilon)=2$.
The fact that the dimension is $2$  reflects the complete 
filling of the allowed area.
The geodesics of the octagon form a self-affine fractal \cite{lb}.
We find the topological entropy, the number and location of fixed points, 
and the spectrum of dimensions in Ref. \cite{lb}.

We suggest the underlying tangle of geodesics could be reflected in the
distribution of large-scale structure \cite{lb}.
The largest structures in the universe have their origin in quantum
fluctuations.  The phenomenon of scarring in quantum chaos along the periodic
orbits \cite{heller} could lead to an enhanced filamentary structure along the
shortest loops through the finite space.
The web of galaxies and clusters of
galaxies and the vibration modes excited by gravitational waves 
on the largest scales could reflect these scars.
The presence of negative spatial curvature provides a natural scale with which
to associate the finite topology. In flat universes, which have been
extensively studied but where chaotic geodesics do not occur, there is no
natural length scale on which to produce topological
identifications and it is entirely ad hoc to create a fundamental
topological identification scale so close to the present Hubble length.
However, if a non-zero cosmological constant exists, as recent
observations of distant supernova may be indicating \cite{saul}, 
then the cosmological
constant provides another fundamental length scale close to the current
Hubble scale with which to associate topological identifications even in a
zero curvature universe. 

In the absence of inflation, there is no dynamical mechanism to generate
large-scale fluctuations.  They are simply an initial condition.  A universe
created finite and hyperbolic can be thought of as a realization from an
ensemble of finite spaces with a spectrum of fluctuations atop a nearly
constant negative curvature manifold.
The spectrum of
fluctuations will then be shaped according to the predictions of quantum chaos.

The tenents of quantum chaos imply that the ordered remnants of
classical chaos
are washed out in the transition to quantum mechanics \cite{{gutz1},{bt},{berry}}.
This expectation is based on two conjectures.  
As suggested by Berry \cite{berry}, the
quantum eigenmodes are well described as concentrated on the region of phase
space traced out by a typical orbit over infinite times.
For a completely chaotic system the orbits cover the entire
space which seems to argue for
a featureless distribution of the quantum modes.
The amplitude of quantum fluctuations are also
conjectured to be drawn from a Gaussian random ensemble with a flat spectrum,
consistent  with the predictions of Random Matrix Theory. 
While these assumptions seem to argue for uniformity in the quantum
fluctuations, they are not inconsistent with striking geometric features. 
Typical eigenstates in a chaotic quantum system have shown scars of enhanced
probability along short period orbits
\cite{heller}.  The scars are consistent with Berry's
conjecture as typical orbits will spend the most time tracing short period
loops.  
The scars can be related to the classical fractal of closed loops.
For a completely chaotic system the fractal will fill the space with a
box counting dimension equal to the dimension of the space, as we
found to be the case for the compact octagon.  
However, if regions
of the fractal are visited more frequently than others, as the shortest
closed loops are in a compact space, 
then the scars might result \cite{lb}.

Scars can be regions of underdensity as well as overdensity.
The consequence for the build up of structure on the largest scales
could be tendency to align with some short period orbits.
in a $2d$ universe, we might see a web-like distribution of clusters
aligned along the orbits of fig. \ref{firstorb}, while structure on 
smaller scales would look featureless.

The scars would have little effect on the 
cosmic microwave background (CMB) although
evidence of topology will be conspicuous through patterns or correlated
circles
\cite{{css},{lbbs},{bps}}.  
The surface of last scattering is also not likely to cut right through a
scar.  As a result, it is reasonable that the CMB will appear smooth when the
distribution of galaxies does not.
The galaxies might be marking the path of the short period orbits,
providing a map of the shortest route around a finite cosmos.

\vskip 15truept

JDB is supported by a PPARC Senior Fellowship. JL is supported by PPARC.

\end{document}